\documentclass[american,british,english]{elsarticle}
\usepackage[LGR,T1]{fontenc}
\usepackage[latin9]{inputenc}
\usepackage{textcomp}
\usepackage{amsmath}
\usepackage{graphicx}

\makeatletter

\DeclareRobustCommand{\greektext}{%
  \fontencoding{LGR}\selectfont\def\encodingdefault{LGR}}
\DeclareRobustCommand{\textgreek}[1]{\leavevmode{\greektext #1}}
\ProvideTextCommand{\~}{LGR}[1]{\char126#1}

\providecommand{\tabularnewline}{\\}

\makeatother

\usepackage{babel}
\begin{document}

\begin{frontmatter}{}

\title{Evaluation of spike sorting algorithms: Simulations and application
to human Subthalamic Nucleus recordings}

\author{Jeyathevy Sukiban$^{*1,2}$, Nicole Voges$^{*2}$\let\thefootnote\relax\footnote{\textit{Email address}: n.voges@fz-juelich.de}
\addtocounter{footnote}{-1}\let\thefootnote\svthefootnote, Till A. Dembek$^{1}$,
Robin Pauli$^{2}$, Michael Denker$^{2}$, Immo Weber$^{3}$, Lars
Timmermann$^{1,3}$, Sonja Gr\"un$^{2,4}$}

\address{$^{1}$ Department of Neurology, University Hospital Cologne, Germany }

\address{$^{2}$ Institute of Neuroscience and Medicine (INM-6) and Institute
for Advanced Simulation (IAS-6) and JARA BRAIN Institute I (INM-10),
J\"ulich Research Centre, Germany }

\address{$^{3}$ Department of Neurology, University Hospital Giessen \& Marburg,
Marburg, Germany}

\address{$^{4}$ Theoretical Systems Neurobiology, RWTH Aachen University,
Germany}

\begin{abstract}
An important prerequisite for the analysis of spike synchrony in extracellular
recordings is the extraction of single unit activity from the recorded
multi unit signal. To identify single units (SUs), potential spikes
are first extracted and then separated with respect to their potential
neuronal origins ('spike sorting'). However, different sorting algorithms
yield inconsistent unit assignments which seriously influences the
subsequent analyses of the spiking activity. 

To evaluate the quality of spike sortings performed with different
algorithms (K-Means, Valley-Seeking, Expectation Maximization) offered by a standard commercial
spike sorter ('Plexon Offline Sorter') we first apply these algorithms
to experimental data (ED), namely recordings in the Subthalamic Nucleus
of patients with Parkinson\textquoteright s disease, obtained during
Deep Brain Stimulation surgery. Since this procedure leaves us unsure
about the best sorting method for the ED we then apply all methods again to artificial
data (AD) with known ground truth (GT). AD consists of pairs of SUs
and perturbation signals embedded in the background noise adapted
to mimic the ED. We generate four AD sets which differ in the similarity
of embedded SU shapes. 

The spike sorting evaluation is performed in terms of the influence
of the respective methods on the SU assignments (e.g., number of units)
and its effect on the resulting firing characteristics (e.g., firing
rates). We find a high variability in the sorting results obtained
by different algorithms that increases with SU shape similarity. We
also find significant differences in the resulting firing characteristics
of the ED. We conclude that Valley-Seeking produces the most accurate
results if the identification of perturbation signals (i.e., artifacts)
as unsorted events is important. If the latter is less important ('clean'
data) K-Means is a better option. Our results strongly argue for the
need of standardized validation procedures for spike sorting based
on GT data. The recipe suggested here is simple enough to become a
standard procedure. It provides a good basis for the evaluation of
spike sorting results in order to ensure reliability and reproducibility. 
\end{abstract}

\end{frontmatter}{}

\tableofcontents{}

\newpage{}

\section{Introduction\label{intro}}

The decomposition of extra-cellular multi unit recordings into single
unit (SU) activity is a prerequisite for studying neuronal activity
patterns \citep{Chibirova2005,Einevoll2012,Kuhn2005,Todo2014,Yang2014}.
The assignment of spikes to individual neurons based on the similarity
of their spike shapes is a method commonly referred to as spike sorting
\citep{Quiroga2007,Lewicki1998,Quiroga2012}. It is composed of three
principal steps. First, the spikes are extracted from the high-pass
filtered raw signal. Second, salient features of each spike waveform
are identified. A common method to automatically extract such features
is the principal component analysis \citep{Lewicki1998,Quiroga2007}.
Third, a sorting algorithm assigns spikes to putative single neuronal
units using the extracted features. Many such spike sorting algorithms
are available \citep{Lewicki1998,Quiroga2007,Quiroga2012,Rossant2016,Chung2017,yger2018}
but they typically provide inconsistent results for the same data
set \citep{Brown2004,Knieling2016,Wild2012}. Such differences in
the sorting results affect the subsequent spike train analyses \citep{Brown2004,Pazienti2006,Todo2014}.
Therefore, a major challenge is to identify an appropriate spike sorting
algorithm for a given data set, considering also its impact on the
subsequent analysis \citep{Lewicki1998,Todo2014}. 

Extracellular recordings from the Subthalamic Nucleus (STN) of patients
with Parkinson's Disease (PD), obtained during Deep Brain Stimulation
(DBS) surgery provide important information about pathological activity
patterns. The analysis of the corresponding SU activity contributes
to identify and localize pathological patterns, e.g., helping to improve
the positioning of the stimulation electrode \citep{Hutchison1998}.

Besides common spike sorting problems such as bursting activities
and overlapping spikes \citep{Lewicki1998,Quiroga2007}, or waveform
changes induced by an electrode drift, the separation of SUs in human
STN data is particularly challenging \citep{Knieling2016}. Microelectrode
recordings from brain areas densely packed with neurons, such as the
STN \citep{Hamani2004}, contain spikes from a large number of neurons.
The overall recording time is restricted to only a few minutes per
recording site since the surgery is exhausting and the patients have
to stay awake\footnote{The aim of the DBS procedure is not the recording itself but to locate
the optimal stimulation site}. The short recording time does not allow to wait for stabilization
of tissue and electrode. In contrast, animal studies allow for longer
recordings so that it is possible to account for initial stabilization
\citep{Raz2000}. Also, simultaneous intra- and extracellular recordings
for calibration can be performed in animal studies \citep{Harris2000}
but this is not feasible during DBS surgery. Another advantage in
animal studies is the usage of 4-wire close-by electrodes (i.e., tetrodes)
or even polytrodes \citep{Rossant2016,Rey2015}. The resulting recordings
generally enable a more accurate spike sorting because one neuron
is registered at different wires allowing for triangulation \citep{Harris2000,Rossant2016,yger2018,Lef2016,Buzsaki2004}.
In contrast, human DBS recordings are typically performed with up
to five single-wire electrodes \citep{McNaughton1983}, typically
inserted using a \textquotedblleft Ben-gun\textquotedblright{} configuration
\citep{Florin2008,Reck2012,Gross2006,Michmizos2010}. These electrodes
have a maximum diameter of 1$\,$mm with a distance of 2$\,$mm. Thus,
the insertion causes a considerable initial tissue movement and the
spikes of one neuron are detected on one electrode only. 

A few comparative spike sorting frameworks for human STN recordings
have been introduced \citep{Chibirova2005,Knieling2016,Wild2012}.
Wild et al. \citep{Wild2012} compares three widely used open source
sorting toolboxes (WaveClus \citep{Quiroga2004}, KlustaKwik \citep{Harris2000},
and OSort \citep{Rutishauser2006}) by applying them to artificial
data with some STN characteristics. They conclude that WaveClus yields
the best results, but does not perform optimally. Knieling et al.
\citep{Knieling2016} compares a new approach to sort STN spikes to
Osort and WaveClus, using the artifical data from \citep{Wild2012}.
However, to our knowledge, there is no comparative study using both
artificial data and STN recordings.

The above studies concentrate on open source spike sorting algorithms,
whereas many studies recording from the human STN \citep{Moran2008,Shinomoto2003,Yang2014,Schrock2009,Kelley2018,Lipski2018,Shimamoto2013}
use a commercially available software, the 'Plexon Offline Sorter'
OFS. Because of its frequent usage and relevance in the scientific
community we focus our studies on the various sorting algorithms offered
by the OFS (see Sec.$\,$\ref{SS}). There are some comparative studies
for spike detection and feature extraction \citep{Lewicki1998,Adamos2008,Gibson2008,Yang2011,Wheeler1982},
but less studies focus on clustering. Here, we concentrate on the
comparison of the results obtained with the following cluster algorithms:
Template Matching (TMS), K-Means (KM), Valley Seeking (VS), standard
and t-distribution Expectation Maximization (StEM and TDEM, respectively).
Varying the cluster algorithm, we use an identical detection procedure
and the first two or three principal components (PCs) as features,
since the number of PCs to be used for the sorting is another matter
of debate \citep{Hulata2002}.

Firstly, we apply all sorting algorithms to the experimental STN data
(ED) recorded from PD patients. This enables us to depict the method-dependent
differences in the ED sorting results (see Fig.$\,\ref{figSort}$)
and to subsequently point out their considerable impact on the analysis
of spike trains from real-world recordings. The evaluation of SU assignments
and properties yield significant differences in the spike sorting
results and suggests a seemingly best method. For a quantitative comparison,
however, ground truth (GT) data are necessary, i.e., data with known
SU assignments \citep{Kretzberg2009,yger2018,Wild2012}. We therefore
generate artificial data (AD) with known GT that include several features
that are close to those of STN recordings. The spike sorting algorithms
are then applied to the AD to evaluate their sorting quality. Based
on this procedure, we are finally able to identify which methods work
best under which circumstances. 

\section{Material and Methods\label{MM}}

We first briefly describe the ED, followed by a description of the
AD generation. Then, we explain the main steps of spike sorting and
finally, we detail the comparison and validation of the results of
different clustering algorithms. 

\subsection{Experimental Data\label{ED}}

ED were recorded intraoperatively from six awake patients with tremor-dominant
idiopathic PD undergoing STN-DBS surgery. The STN was localized anatomically
with preoperative imaging and its borders were intraoperatively verified
by inspection of multi unit spiking activity. Up to five combined
micro-macro-electrodes recorded single cell activities and LFPs using
the INOMED ISIS MER-system 2.4 beta, INOMED Corp., Teningen, Germany.
Four of the electrodes were distributed equally on a circle with 2$\,$mm
distance from the central electrode using a Ben Gun electrode guide
tool. ED were already analyzed in a previous study \citep{Florin2012}
which was approved by the local ethic committee. For more detailed
information about the recording setup and recording procedures see
\citep{Florin2008,Reck2012,Gross2006,Michmizos2010}.

The microelectrodes had an impedance of around 1M\textgreek{W} during
each recording session. The signal was amplified by a factor of 20$\,$000,
band-pass filtered from 220 to 3000$\,$Hz, using a hardware Bessel
filter, and sampled at 25$\,$kHz by a 12$\,$bit A/D converter (+/-
0.2$\,$V input range). Recording started 6$\,$mm above the previously
planned target point. The extracellular multi unit signals were recorded
after moving the electrode closer to its target in 1$\,$mm steps. 

\begin{figure}
\includegraphics[width=1\columnwidth]{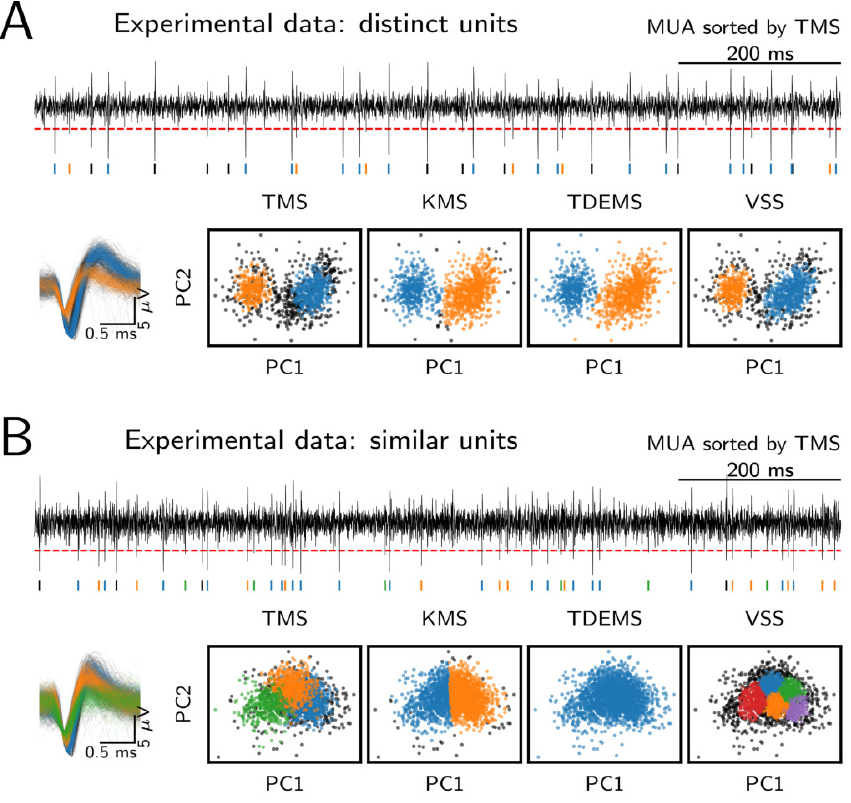}\caption{\textbf{Exemplary spike sorting of human STN recordings. }\label{figSort}
(A) Two distinct SUs and (B) several similar SUs that were extracted
from two STN recordings (top traces in A and B) by amplitude thresholding
(horizontal red line). Colored vertical lines below the continuous
traces indicate time stamps of potential spikes. Extracted spike waveforms
are shown on the left, the corresponding clustering in feature space
on the right: colored dots represent spikes, US in gray. }
\end{figure}

A total of 38 STN recoding traces from six awake PD patients at rest
with one to four simultaneous microelectrode trajectories in different
recording depths were analyzed. Some example sortings are shown in
Fig.$\,$\ref{figSort}. The inclusion criteria for a data trace were
a) a minimum length of 20$\,$s, b) no drifts in background activity,
c) spiking activity in the STN (based on visual inspection), and d)
not exceeding the dynamic range of the A/D converter. The longest
stable segment of a given trace was selected for further analysis;
the first 2$\,$s of each recording after electrode movement were
discarded. 

\subsection{Artificial data generation \label{AD}}

A rigorous way to compare spike sorting methods is to test them on
data sets with known GT, i.e., we know which of the spikes originate
from which neuronal units. To this end, we generate AD by first selecting
the two most distinct average spike \foreignlanguage{american}{wave}\foreignlanguage{british}{forms}
from one ED trace. To enhance their differences, the larger one is
multiplied with a factor of 1.1 so that it exhibits more pronounced
peak amplitudes than given in the ED\footnote{The smaller one is kept as it is}.
We call the waveforms w1 and w2. We then linearly combine \foreignlanguage{american}{w1
and w2} to obtain spike pairs (u1,u2) whose similarity can be varied
parametrically: 

\begin{align}
\text{u1} & =\lambda\qquad\;\;\cdot\text{w1}+\:(1-\lambda)\cdot\text{w2}\\
\text{u2} & =(1-\lambda)\cdot\text{w1}+\:\lambda\qquad\;\;\cdot\text{w2 \ensuremath{\quad\qquad}with \ensuremath{\lambda\in}\ensuremath{[0.5,1]}}\nonumber 
\end{align}

\begin{figure}
\includegraphics[width=0.66\columnwidth]{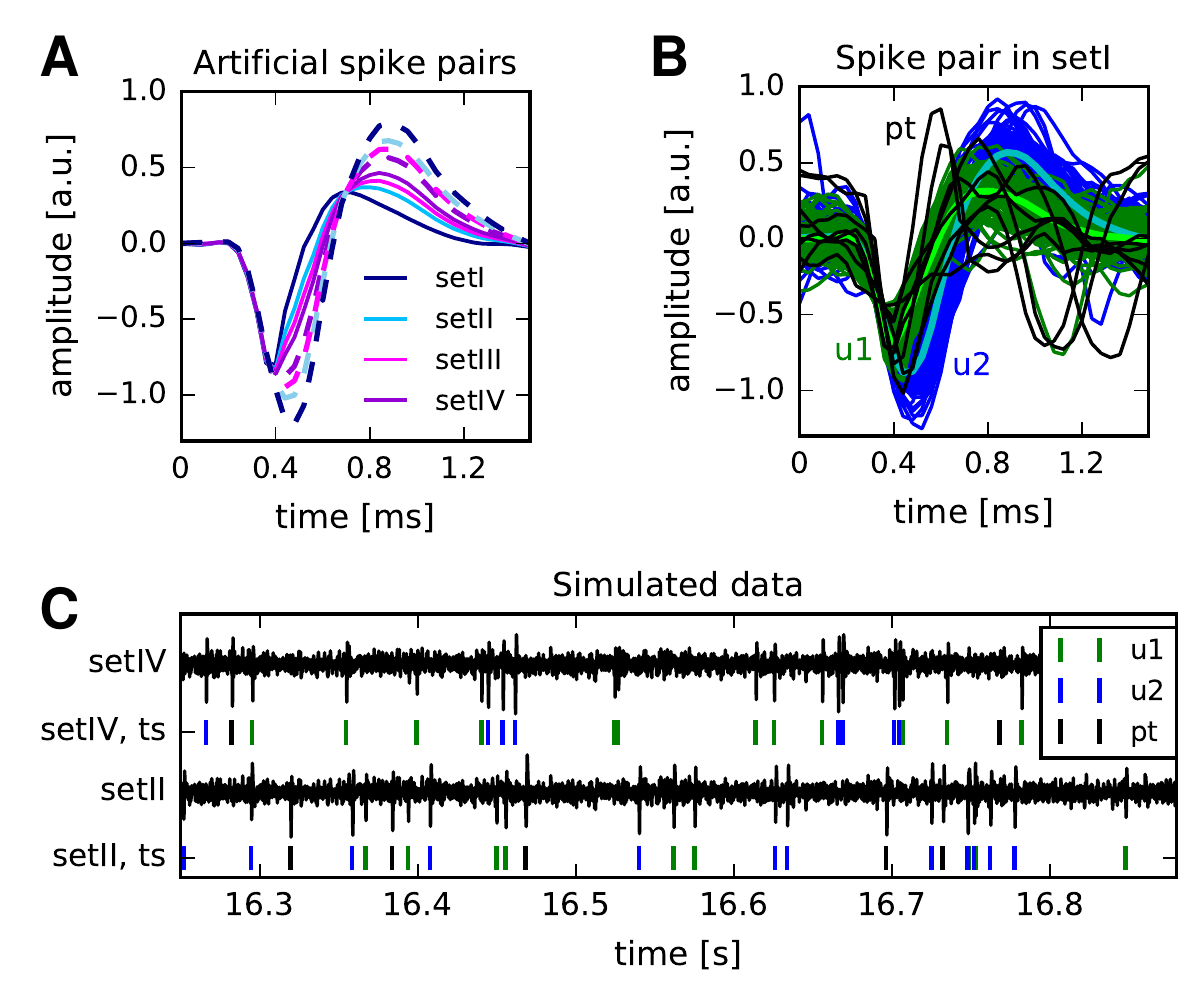}\caption{\textbf{Artificial data generation. }\label{FigADgen} (A) Four pairs
(solid and dashed curves) of artificial spikes that are very distinct
(setI, blue), distinct (setII, cyan), similar (setIII, magenta), and
very similar (setIV, violet). (B) Exemplary setI spike pair (u1,u2)
added to the noise (100 spikes each). Thick light blue and green lines
indicate the average spike waveforms of u1 and u2, respectively, black
lines indicate perturbations (pt). (C) Simulated recording traces
(black lines) with the corresponding ground truth spike times: Each
trace contains one particular (u1,u2) pair as well as pts (green,
blue and black markers, respectively).}
\end{figure}

Thus, by varying $\lambda$ we create data sets with different degrees
of similarity of the spike pairs (u1,u2), see Fig.$\,$\ref{FigADgen}A.
For $\lambda=1$ we obtain u1=w1 and u2=w2 with u1 and u2 being most
different. For $\lambda=0.5$, u1 and u2 are identical. We generate
four AD sets, each with one spike pair (u1,u2) obtained for a certain
value of $\lambda$ with $\lambda=1,0.8,0.7,0.6$. The corresponding
data sets are called setI ($\lambda=1$, most distinct pair), setII,
setIII and setIV ($\lambda=0.6,$ most similar pair). The hypothesis
behind this choice is that it should be easier to distinguish distinct
spikes than similar spikes. 

The spike pairs are then added to background noise (Fig.$\,$\ref{FigADgen}B,C).
To obtain the noise as realistic as possible, we generate it from
the ED, using concatenated recording intervals without any spikes.
We reshuffle the phase of the original noise so that the power spectrum
is kept constant. The respective pairs of spikes (u1,u2) are added
to the noise, each at a rate of 14$\,$Hz as estimated from the ED,
assuming a Poisson distribution. Each of the four generated data sets
has a length of 40$\,$s, sampled at 25$\,$kHz resulting in approx.$\,$500
spikes per unit. Refractory period violations (rpv, see Sec.$\,$\ref{EvalQ})
are corrected for by shifting the corresponding spikes of each SU
in time (the second one is shifted forward by 1 to 50 time stamps
depending on the closeness of the two spikes) until no more rpv are
found. 

Inspired by the difficulties that occur during sorting the ED, we
additionally include the following challenges: We inject overlapping
spikes from different SUs by inserting u1 spikes 10 to 22 time stamps
(randomly chosen) after some randomly chosen u2 spikes (approx.$\,$2.5\%
of the total number of spikes per trace as estimated from the ED)
and vice versa. We then again correct for rpv. Moreover, we add in
total approx.$\,$100 so-called perturbation (pt) signals to each
trace. These represent artifacts, e.g., noise originating from electrical
equipment which may resemble spikes \citep{Horton2007}. Perturbations
are given by 8 sinusoidal functions (black lines in Fig.$\,$\ref{FigADgen}B).
Each pt consists of one cycle of the following frequencies $f=1,0.75,0.5,0.25$
with respect to $T=1.52$$\,$ms spike length, using a positive amplitude
of either the peak amplitude of u2 or half of it. The negative peak
amplitude is fixed to the minimum of the spike generated with $\lambda=0.5$.
The corresponding insertion times are again Poisson distributed with
a firing rate of 3.3$\,$Hz as estimated from the ED. The aim is to
investigate how different sorting methods deal with such perturbations.
Ideally, pt signals should be left as unsorted events (US). 

We generate 10 realizations for each of the four AD data sets. The
noise is identical for the four data sets within one realization (setI
to setIV) but changes between the 10 realizations. After inserting
potential spike events, i.e., u1, u2 and pts, into the noise, some
threshold crossings vanish while some new crossings without a corresponding
AD event emerge, e.g., due to possible overlaps between u1, u2 and
pt. Therefore, the spike times of the GT are obtained as follows:
(1) Calculation of the spike detection threshold, i.e., mean signal
minus four times the standard deviation (SD) of the complete trace,
identical for all sorting methods and identical to the procedure used
for the ED (cf.$\,$Sec.$\,$\ref{SS}). (2) Insertion of u1 spikes
into the pure noise and detection of threshold crossings. (3) Repetition
of step (2) for u2 and pt signals, respectively. When comparing the
spike times in GT and sorting results we only consider threshold crossings
that occurred after insertion and that have a corresponding AD event
time stamp. We allow for deviations up to +/- 0.5$\,$ms. AD were
generated using Python2.7.

\subsection{Spike detection and spike sorting \label{SS}}

ED and AD are separated into SUs using spike sorting algorithms implemented
in the 'Plexon Offline Sorter' OFS (Offline SorterTM, Plexon Inc.,
Dallas, TX, USA). During a pre-processing step, artifacts (i.e., non-physiological
events that may resemble spikes, e.g., some of the pts in the AD)
were identified by visual inspection and removed. The spike detection
threshold was set to minus four SD of the background noise\footnote{Exceptionally we also used 4.5 SD, depending on the individual signal-to-noise-ratio
of the ED spikes. } \citep{Mrakic-Sposta2008}. After detection, 360$\,$\textmu s before
and 1160$\,$\textmu s after threshold crossing were extracted, resulting
in a total spike length of 1520$\,$\textmu s (38 time stamps). The
spikes are aligned at the point of threshold crossing (see Fig.$\,$\ref{FigADgen}B). 

Several features of the waveforms such as peak and valley amplitude,
peak-valley distance, energy of the signal, and PCs were extracted.
For the supervised 'manual' sorting method TMS (see Sec.$\,$\ref{TMS}),
all extracted features were used to visually identify the templates
and thus the number of clusters, while the clustering itself uses
the complete waveforms. For all other algorithms, clustering is solely
based on the first two (2D) or three (3D) principal components. We
apply the sorting algorithms TMS, KM, VS, StEM, and TDEM (the methods
are described in the following subsections) to both AD and ED.

\begin{figure}
\includegraphics[width=0.5\columnwidth]{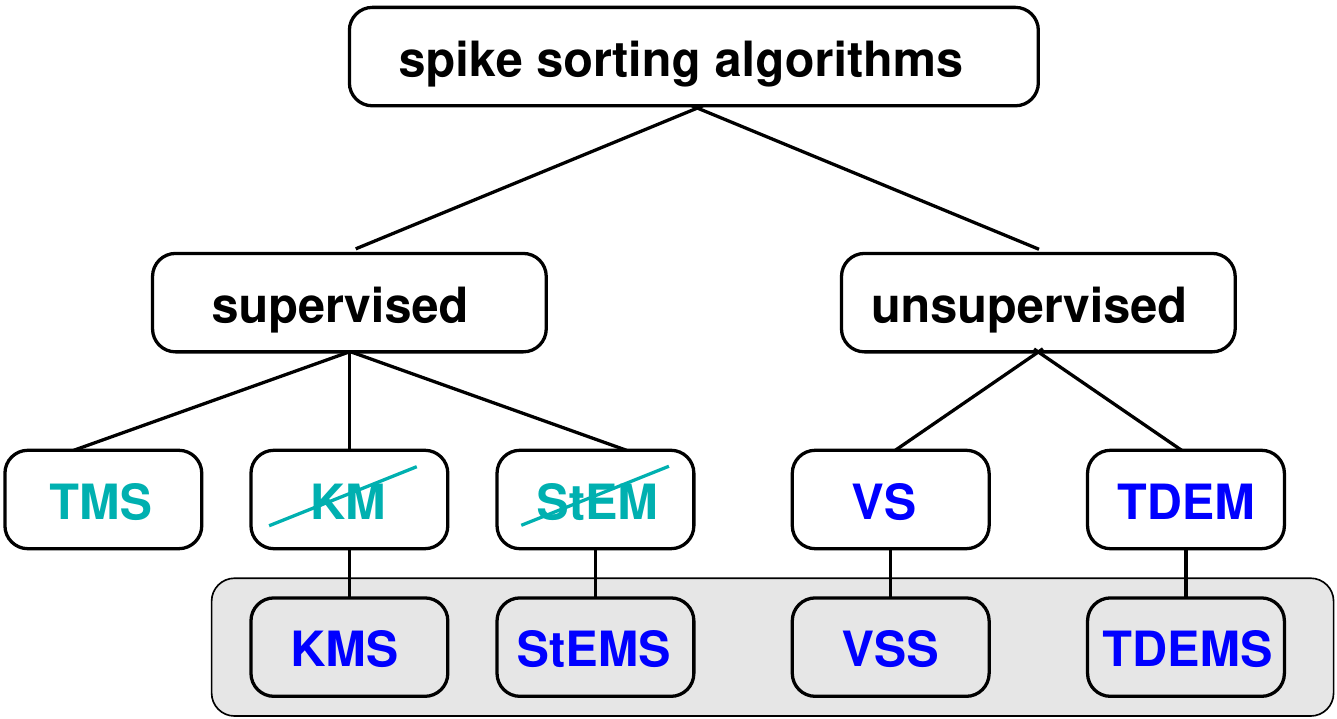}\caption{\textbf{Spike sorting algorithms.} \label{figAlg}Schematic overview
of different sorting methods separated into originally supervised
(cyan) and unsupervised algorithms (blue). Algorithms that are crossed
out were only used in combination with a parameter scan (gray background)
to enable unsupervised clustering.}
\end{figure}

VS and TDEM (see Secs.$\,$\ref{VSS} and$\,$\ref{EM}) automatically
determine the number of resulting clusters (unsupervised clustering)
but contain method-specific parameters which were set to default values
d (see Tab.$\,$\ref{tab1}). In addition, these methods were applied
in combination with a parameter scan which optimizes the method-specific
parameters (called VSS, TDEMS if used with a scan). During such a
scan, a spike sorting algorithm runs repetitively for a wide range
of parameter values (varied by step size $\Delta$) to identify and
select the run that yields the best sorting quality based on cluster
quality metrics, e.g., distances in feature space. TMS, KM, and StEM
require user intervention. To enable unsupervised clustering, KM and
StEM are only applied in combination with a parameter scan (KMS, StEMS)
which automatically computes the appropriate number of clusters. Table$\,$\ref{tab1}
and Fig.$\,$\ref{figAlg} list the methods that are used with a scan,
as well as the corresponding parameter ranges.

\begin{table}
\caption{Methods used with a scan and their tunable parameters. The last column
indicates the range of tested values, the step size $\Delta$, and
the default value d if used without scan. \label{tab1}}

\begin{tabular}{|c|c|c|}
\hline 
Method & Scanning parameter & Scanning range\tabularnewline
\hline 
\hline 
KMS & \#SUs & 1-7 SUs\tabularnewline
\hline 
StEMS & \#SUs & 1-7 SUs\tabularnewline
\hline 
TDEM(S) & DOF multiplier & 10 to 30, $\Delta$=5, d=10\tabularnewline
\hline 
VS(S) & Parzen multiplier  & 0.5 to 1.5, $\Delta$=0.2, d=1\tabularnewline
\hline 
\end{tabular}
\end{table}

In total, 13 different sorting approaches were applied to each data
trace (TMS, VSS2(3)D, VS2(3)D, KMS2(3)D, StDEMS2(3)D, TDEMS2(3)D,
TDEM2(3)D, cf. Fig.$\,$\ref{figAlg}). We apply each method using
the first two (2D) or three (3D) PCs, enabling a comparison of the
corresponding performances. To investigate the effect of using a parameter
scan, we apply each method with a method-specific parameter twice,
once with parameter scan and once using its default value (see Tab.$\,$\ref{tab1}).
After spike sorting, each threshold crossing event was either labeled
as sorted into a cluster (SU) or left unsorted (US) if no clear assignment
could be made. In the following subsections, we give more details
on the spike sorting algorithms used in our study. 

\subsubsection{Template Matching sorting (TMS)\label{TMS}}

TMS is a supervised clustering algorithm, the number of clusters has
to be predefined by the user. Based on various features (cf.$\,$Sec.$\,$\ref{SS})
the user selects one waveform as template for each cluster. Then,
the algorithm calculates the root-mean-square differences $D_{w}$
for all waveforms $w$ to these templates $t$: $D_{w}=\sqrt{1/N{\displaystyle {\textstyle \sum_{i=1}^{{\scriptstyle N}}(w_{i}-t_{i})^{2}}}}$,
where $N$ is the number of time stamps per waveform. TMS identifies
the template with minimum difference $D_{w}$ for each single waveform.
If the minimum is smaller than a user defined value for the allowed
variability, the particular waveform will be assigned to the cluster
defined by this template. 

\subsubsection{K-Means clustering as Scan (KMS) \label{KMS}}

The K-Means algorithm requires the user to select a predefined number
of clusters and the corresponding cluster centers which are here provided
by the scanning algorithm. First, each sample point, i.e., each waveform
in PC feature space is assigned to the nearest cluster center, based
on Euclidean distances. Second, the cluster centers are recalculated
using the center-of-gravity method \citep{Dai2008,Wilkin}. Steps
one and two are repeated until convergence is reached, i.e., clusters
centers are stable. Finally, outliers are removed: Based on mean ($\mu$)
and SD of the distances of all sample points from their cluster center,
a sample point is removed if it exceeds the outlier threshold, set
to $\mu+2\cdot$SD. It is then left as US.

\subsubsection{Valley Seeking (VS) \label{VSS}}

The VS algorithm is based on an iterative non-parametric density estimation
\citep{Fukunaga1990,Zhang2007}. To subdivide the sample points (i.e.,
spikes) into exclusive clusters, the algorithm estimates their density
in PC feature space using the Parzen approach \citep{Fukunaga1990},
which estimates the appropriate kernel and its width for the best
separation. VS calculates the number of neighbors of each sample point
in a fixed local volume and determines the valleys in the density
landscape. The critical radius R of the fixed volume is defined as
$\text{{R}}=0.25\cdot\sigma\cdot\text{{PM}}$, where $\sigma$ is
the SD of the distances of all samples to the overall center point,
and PM is the Parzen multiplier, a user-defined parameter. A sample
point becomes a seed point of a cluster if its number of neighbors
exceeds a threshold. Then, initial clusters are formed by the seed
points with the most neighbors. An iterative process classifies still
unassigned sample points or leaves them unsorted. We run VS both with
the PM default value, and using the scanning algorithm for PM (VSS).

\subsubsection{Expectation Maximization algorithms (EM) \label{EM}}

The standard EM (StEM) algorithm is an iterative, parametric approach
that assumes that several Gaussian distributions underlie the distribution
of sample points (i.e., spikes). The algorithm requires the user to
select the number of Gaussians to be fitted and to define the initial
cluster centers \citep{Fukunaga1990,Sahani1999}. To enable unsupervised
clustering these are provided by the scanning algorithm. The algorithm
starts by running the K-Means algorithm for the first assignment of
sample points from which the initial Gaussian parameters are estimated.
An iterative process optimizes these parameters until convergence
of the Gaussian distributions to stable values. Each iteration consists
of an expectation (E)-step that calculates the likelihood for each
sample point to belong to each Gaussian, and a maximization (M)-step
that maximizes the expected likelihood by optimizing the parameters
\citep{Fukunaga1990,Sahani1999}. 

The t-distribution EM-algorithm (TDEM) differs from the StEM by fitting
wide-tailed t-distributions instead of Gaussians. It has been shown
that t-distributions yield a better fit to the underlying statistics
of the waveform samples \citep{Shoham2003}. TDEM directly provides
unsupervised clustering by starting with a large number of clusters
and iteratively optimizing the likelihood function (assignment of
samples to clusters) \citep{Shoham2003}. The shape of the t-distribution
is determined by the degree of freedom (DOF) multiplier which depends
on the sample size and controls the convergence properties \citep{Figueiredo2002,Shoham2003}.
We run TDEM both with the DOF default value, and using the automatic
scanning algorithm for DOF. In the Plexon implementation of these
EM algorithms no events are left unsorted.

\subsection{Evaluation of spike sorting results \label{Eval}}

Sorting results are characterized by the number of detected SUs and
the number of unsorted events (US). The resulting means and SDs per
data set are calculated by averaging over the 10 realizations for
each AD set and the 38 ED recording traces, respectively. 

\subsubsection{Comparison with ground truth data\label{EvalGT}}

To evaluate the accuracy of the clustering algorithms, the resulting
SUs were compared with a given GT. To quantify accordance with and
deviations from the GT, we calculate the following numbers (cf.$\,$Fig.$\,$\ref{figED2}D
and Fig.$\,$\ref{figGT}D): 
\begin{description}
\item [{TP}] true positives, i.e., correctly assigned spikes: a waveform
was given as element of a certain SU and was sorted into this SU.
\item [{FP}] false positives (sorted), i.e., wrongly assigned (misclassified)
spikes: a waveform was given as element of a certain SU but was sorted
into another SU.
\item [{FN}] false negatives, i.e., spikes wrongly left unsorted: a waveform
was given as element of a certain SU but was left unsorted.
\item [{FPp}] false positives (unsorted), i.e., wrongly assigned (misclassified)
perturbations: a pt signal was classified as element of a certain
SU.
\item [{TN}] true negatives, i.e., correctly assigned perturbations: a
pt signal was left unsorted.
\end{description}
Thus, a 100$\,$\% correct classification contains only TP and TN.
For each data set sorted by a certain method, we count the corresponding
hits (TP, TN) and misses (FP, FPp, FN) and normalize by the number
of all events (spikes and pts) that are present in both GT and the
corresponding sorting outcome. For each SU of the GT, we check which
unit of the spike sorting outcome contains the most hits and then
take this unit as correct. Therefore, we always find TP $>$ FP. Based
on these numbers we calculate the following measures. The sensitivity
describes how many spikes out of all spike events are correctly assigned:
sensitivity = TP/(TP + FP + FN) while the specificity describes how
many of the pts are correctly left unsorted: specificity = TN/(TN
+ FPp).

These analyses were performed using MATLAB (Mathworks Inc., Natick,
U.S.A.). Differences in the general performance of the algorithms
were evaluated by comparison to the GT values using the Wilcoxon rank
sum test. Bonferroni's correction was applied to adjust the significance
level for multiple comparisons. To contrast the 2D with the 3D version
of a method, we used direct comparisons (Wilcoxon rank sum test without
Bonferroni's correction), as well as for the comparison of running
a method with a scan versus using the default parameter value.

\subsubsection{Quality of spike sorting \label{EvalQ}}

We also assess the quality of our sorting results with the following
evaluation measures: the percentage of refractory period violations
(rpv), the isolation score (IS), and a measure characterizing the
internal cluster distance (Di) \citep{Fee1996,Hill2011,Joshua2007,Einevoll2012}.
The amount of rpv indicates the degree of multi unit contamination
in a given SU. We count the number of inter-spike intervals (ISIs)
smaller than 2$\,$ms divided by the total number of ISIs in this
SU times 100. The IS compares the waveforms within one SU to all other
potential spikes in the recording trace based on the normalized and
scaled Euclidean distances of their time courses \citep{Joshua2007}.
It provides an estimate of how well a SU is separated from all other
potential spikes outside its cluster\footnote{AD were used to calibrate the IS scaling parameter to five.}:
IS=1 means well separated while IS=0 indicates overlapping clusters.
It thus requires the existence of potential spikes outside a given
cluster. Since EM methods do not account for US, we only calculate
the IS when there are at least two SUs in a given trace. The internal
cluster distance Di is also calculated if there is only one SU. This
measure uses the first three PCs of each waveform. For each SU, we
calculate the mean waveform and its mean euclidean distance (in reduced
PC space) to all other spikes inside this cluster. For a consistent
scaling behavior of the latter two quality measures we consider 1-Di
so that high IS and high 1-Di values indicate well defined clusters. 

\subsubsection{Firing properties\label{F} }

To investigate the differences in the dynamical properties of the
SUs we calculate the mean firing rate and the local coefficient of
variation LV \citep{Shinomoto2003} in the ED. The LV characterizes
the firing regularity of a SU:

\begin{equation}
LV=1/(n-1)\sum_{i=1}^{n-1}3(T_{i}-T_{i+1})^{2}/(T_{i}+T_{i+1})^{2},
\end{equation}
where $T_{i}$ is the duration of the ith ISI and n the number of
ISIs. LV values enable the following classification \citep{Shinomoto2003}:
regular spiking for LV$\in[0,0.5]$, irregular for LV$\in]0.5,1]$,
and bursty spiking for LV>1. For this analysis, only units with more
than 80 spikes were taken into account to avoid outliers. 

\section{Results\label{Res}}

We first present the evaluation of the results obtained by applying
the 13 sorting algorithms to the STN recordings (ED), followed by
an investigation of the impact of the sortings on the dynamical properties
of the resulting SUs. The ED sorting evaluation leaves us in doubt
about the best method. Therefore, we then evaluate the results of
applying the identical methods to the AD which allows for an objective
ground truth (GT) comparison. This procedure enables us to finally
identify the best sorting methods.

\subsection{Spike sorting of experimental data\label{ResED}}

We aim at sorting the ED under the additional constraint of identifying
an \emph{unsupervised} sorting algorithm that enables a fast, reliable
and reproducible extraction of SUs. Our criteria for a successful
sorting are: (1) all true spikes are detected and (2) non-spikes (i.e.,
artifacts) are not extracted as spikes but left unsorted. For a quantitative
evaluation, we first sort the data using TMS, a manual sorting method
that puts the user in complete control. According to our criteria
it was performed with precise visual inspection aiming for clearly
separated SUs that are free of artifacts and wrongly classified spikes.
In search for an automatic sorting algorithm, we apply the following
unsupervised methods offered by Plexon: VS and TDEM (both applied
with parameter scan and default value, respectively, in 2D and 3D,
respectively), as well as KMS and StEMS (both only applied with scan,
in 2D and 3D, respectively). Some exemplary results are shown in Fig.$\,$\ref{figSort}.
Here, we assume that the TMS sorting represents the GT, because TMS
is a widely used method \citep{Rutishauser2006,Levy2002a,Raz2000,Steigerwald2008},
subjectively often perceived as the best sorting.

\begin{figure}
\includegraphics[width=0.5\columnwidth]{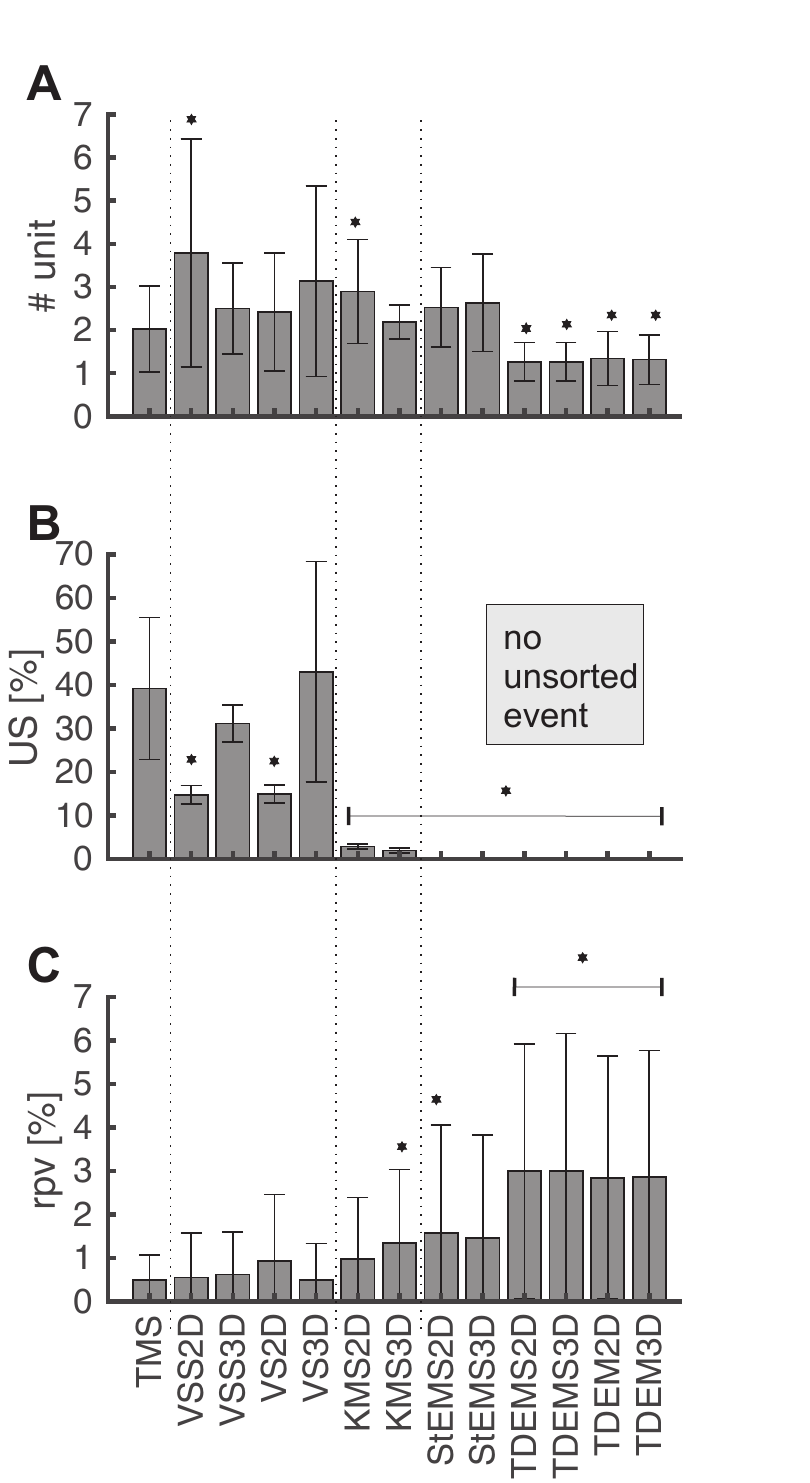}\caption{\textbf{Evaluation of ED sorting results.}\label{figED1} (A, B, C)
Bar plots of the average number of detected SUs (A), the percentage
of US (B) and the percentage of rpv (C), in dependence of the sorting
method. Shown are mean +/- SD of the 38 recording traces. Stars indicate
a significant difference (p<0.0042 after Bonferroni correction) to
the TMS results.}
\end{figure}

For a quantitative analysis we evaluate the number of detected SUs
(Fig.$\,\ref{figED1}$A), the percentage of unsorted events US (Fig.$\,\ref{figED1}$B),
and the percentage of refractory period violations rpv (Fig.$\,\ref{figED1}$C).
The number of detected SUs is highly variable, depending on the sorting
method. TMS and KMS3D detect on average two SUs, TDEM(S) detect on
average significantly less units whereas VSS2(3)D yield significantly
more units than TMS. EM methods do not account for US, they do not
leave any spike unsorted. KMS methods yield the least US, followed
by VS(S)2D while VS(S)3D and TMS show the highest percentage of US.
TMS, all VS methods and KMS2(3)D result in less than 1.5\% rpv. Methods
that do not account for US yield more potential spikes per SU which
results in a higher probability of rpv occurrences. 

\begin{figure}
\includegraphics[width=0.67\columnwidth]{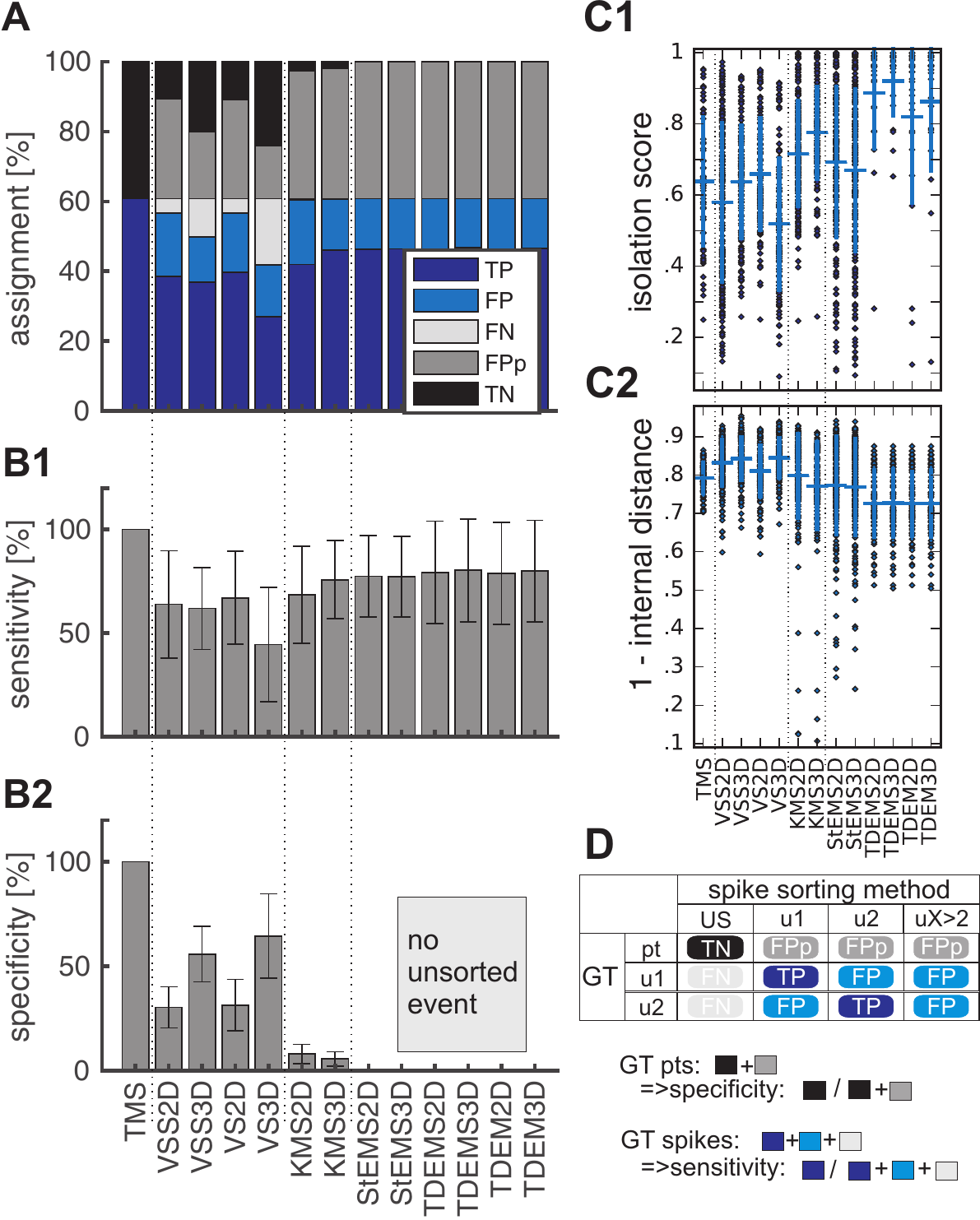}

\caption{\textbf{Evaluation of ED sorting results} with GT (TMS) comparison.
\label{figED2} (A) Stacked bar plot showing the percentage of correct
and wrong assignments of ED spikes in dependence of the sorting methods
using TMS as GT reference: TP indicate correctly classified spikes,
FP misclassified spikes, FN spikes left unsorted, FPp indicate US
events of TMS that are classified as spikes, and TN indicates US events
of TMS that are also left unsorted by the other methods. (B1, B2)
Sensitivity and specificity measure in dependence of the sorting methods.
Shown are mean +/- SD of all recording traces. (C1, C2) Cluster quality
measures IS and (1-Di) applied to ED: each dot represents the value
obtained for one SU in the 38 recording traces. Horizontal lines indicate
the average over all SUs, vertical lines indicate the corresponding
SD. (D) Summary of TP, TN, FP and FN notations and definition of sensitivity
and specificity measures.}
\end{figure}

We now compare the assignments of the different sortings to the GT
given by TMS using the terminology introduced in Sec.$\,\ref{EvalGT}$:
TP, TN, FP, FPp and FN rates (Fig.$\,\ref{figED2}$D). Since TMS aims
at 'clean' SUs it results in a high TN rate of 39\% and thus only
61\% TP, see Fig.$\,\ref{figED2}$A. All other methods leave less
events unsorted, resulting in lower TN (black) and accordingly higher
FPp (dark gray) rates. They show a similar amount of misclassified
spikes (FP, light blue) but clearly differ in terms of their FN (light
gray), FPp and TN rates. EM methods, e.g., yield no TN but only FPp,
since they do not allow for US, hence the high amount of rvp (cf.$\,$Fig.$\,$\ref{figED1}C).
Compared to TMS (the assumed GT), TP rates are reduced for all other
methods, the most for VS methods due to the high FN rates. KMS yield
relatively high TP rates but very low TN rates because only a very
few events are left unsorted (cf.$\,$Fig.$\,$\ref{figED1}B).

The sensitivity (i.e., percentage of TP relative to the total number
of true spikes in TMS) and specificity (percentage of TN relative
to the number of US in TMS) measures in Fig.$\,\ref{figED2}$B1 and
Fig.$\,\ref{figED2}$B2 summarize these results. We aim at both a
high sensitivity (i.e., correctly classified spikes) and a high specificity
(events correctly left unsorted). In total, the sensitivity varies
between 44\% and 80\% (Fig.$\,\ref{figED2}$B1), and the specificity
between 0\% and 64\% (Fig.$\,\ref{figED2}$B2). All EM methods result
in a high sensitivity, but zero specificity, because they do not account
for US. KMS methods also yield a high sensitivity, but a low specificity.
VS methods result in a high specificity combined with a rather low
sensitivity. Combining these two measures, VSS3D seems to provide
the best result, since it shows the highest sensitivity of all VS
methods.

Fig.$\,$\ref{figED2}C1 and Fig.$\,$\ref{figED2}C2 asses the sorting
quality from another perspective, independently of the assumed GT:
The isolation score (IS) and the internal cluster distance (Di) indicate
how well the resulting SUs are clustered (cf.$\:$Fig$\,$\ref{figSort}).
For well separated clusters without artifact contamination, IS and
1-Di are close to one. The large vertical spread indicates a large
variability for all methods, mostly due to the high variability in
the \#SU detected by each method, cf. Fig.$\,$\ref{figED1}A. TMS
yields a rather low IS although the low percentage of rpv indicates
a successful sorting (cf.$\,$Fig.$\,$\ref{figED1}C). The high IS
values for TDEM methods do not indicate well defined clusters due
to a high amount of rpv (cf. Fig.$\,$\ref{figED1}C). They are simply
a consequence of the fact that the IS can only be calculated when
there is more than one SU which is not often the case, cf. Fig.$\,\ref{figED1}$A.
The Di measure considers all SUs and indeed indicates poorly defined
clusters. KMS methods yield relatively high IS and 1-Di values and
a reasonable amount of rpv. VS methods show relatively high 1-Di values
but comparably low IS scores. Together with the high FN rate (Fig.$\,$\ref{figED2}A)
this indicates that many spikes are left unsorted.

In summary, we find that VSS3D agrees best with the TMS results, suggesting
that VSS3D is the best sorting method. However, a detailed comparison
of the assignment of individual spikes indicates considerable differences:
The FPp, FP, and FN rates for VSS3D sum up to approx. 40\%. Other
issues are the low IS score for both TMS and VSS3D, the higher TP
rate for KMS compared to VSS3D, as well as the fact that we might
loose a lot of true spikes when using TMS or VSS3D due to 39\% US.
Moreover, all VS and KMS methods detect more SUs compared to TMS.
Tab.$\,$$\ref{tab2}$ in Sec.$\,$\ref{Appendix} lists more quantitative
details. In the end, we are left with the suspicion that the subjective
TMS sorting and thus VSS3D might, after all, not be the best methods
to sort our data. In Sec.$\,\ref{ResAD}$ we therefore apply all methods
again to AD which provides an objectively given GT to compare with.

\subsection{Impact of spike sorting methods on SU firing properties}

\begin{figure}
\includegraphics[width=1\columnwidth]{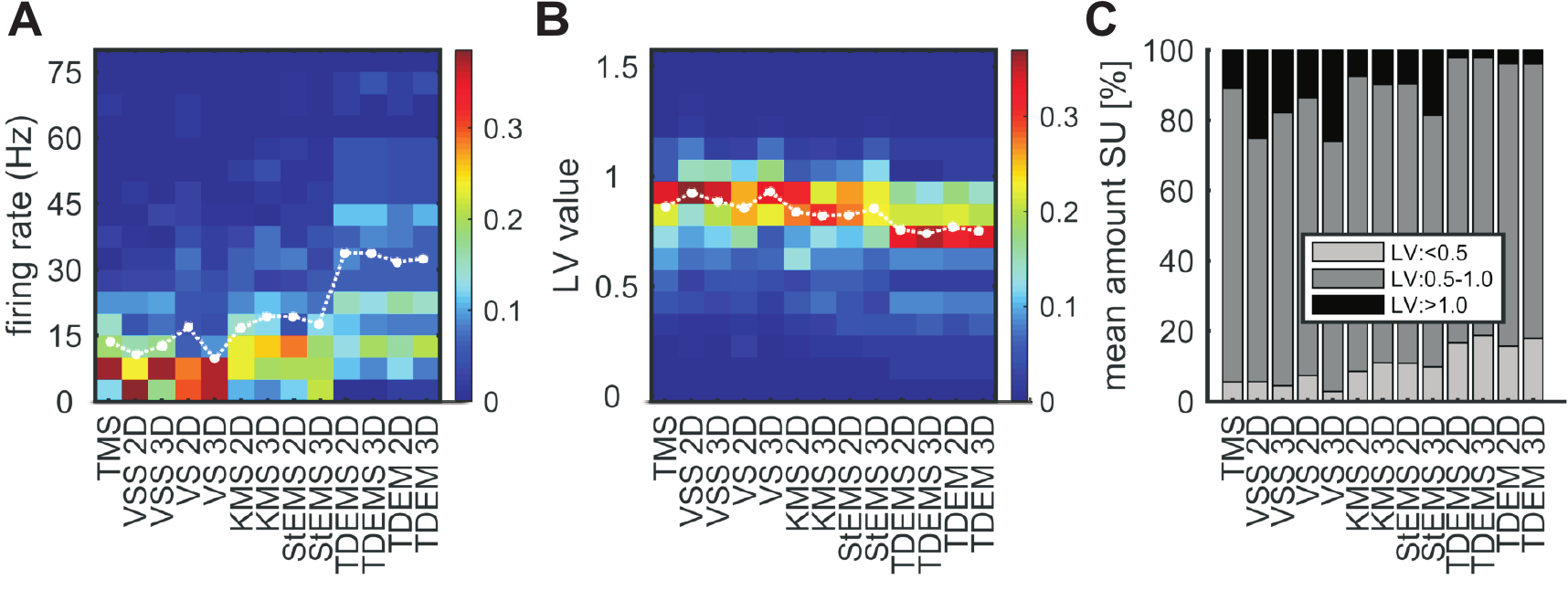}\caption{\textbf{Variability in the ED firing statistics\label{figDyn}.} Differences
in firing patterns characterized by (A) firing rate and (B) local
coefficient of variation LV. The color code indicates the number of
detected SUs normalized to the total number of detected SUs per method
(binned and averaged over all traces). Mean firing rate and mean LV
(averaged over all SUs) are indicated by white dots. (C) Amount of
SUs with regular (LV$<$0.5), irregular (0.5$\protect\leq$LV$\protect\leq$1.0)
and bursty (LV$>$1.0) firing patterns.}
\end{figure}

To characterize the differences in the firing patterns of all detected
SUs that result from using different sorting algorithms, we calculate
the firing rate and the local coefficient of variation for each SU.
Fig.$\,$\ref{figDyn}A and Fig.$\,$\ref{figDyn}B show the corresponding
distributions obtained by binning and averaging across all STN recording
traces. Each entry is averaged over all SUs identified in the ED.
Fig.$\,$\ref{figDyn}A shows clear discrepancies in the firing rate
distributions. This is a consequence of the distinct number of SUs
obtained from the different methods, as well as of the amount of US
(cf. Fig.$\,$\ref{figED1}). SUs obtained with TMS and VS show lower
frequencies (maximally up to 30$\,$Hz), while KMS and StEM yield
SUs with up to 40$\,$Hz. SUs obtained with TDEM methods have the
highest firing rates (60$\,$Hz) since these methods detect mostly
one SU and leave no US. 

Another characteristic property of spiking activity is the LV which
quantifies the regularity in neuronal firing (Fig.$\,$\ref{figDyn}B).
We again observe method-dependent deviations, based on the variable
number of SUs: the lower the number of detected SUs, the more regular
are the subsequent spike trains. When classifying the SUs according
to their LV value in regular (LV<0.5), irregular (0.5$\leq$LV$\leq$1.0),
and bursty (LV>1) firing neurons \citep{Shinomoto2003,Lourens2013}
we find clear differences (\ref{figDyn}C): TDEM methods yield more
regular and less bursty SUs (less than 5\%) whereas VS2(3)D methods
result in less regular and more bursty SUs (up to 25\%).

\subsection{Spike sorting of artificial data\label{ResAD}}

The ED results left us undecided concerning the best sorting method.
In need of an objective GT we now evaluate the results of sorting
artificially generated data, see Sec.$\:\ref{AD}$. For the AD we
know the correct spike and perturbation (representing artifacts to
be left unsorted) assignments.

\begin{figure}
\includegraphics[width=0.67\columnwidth]{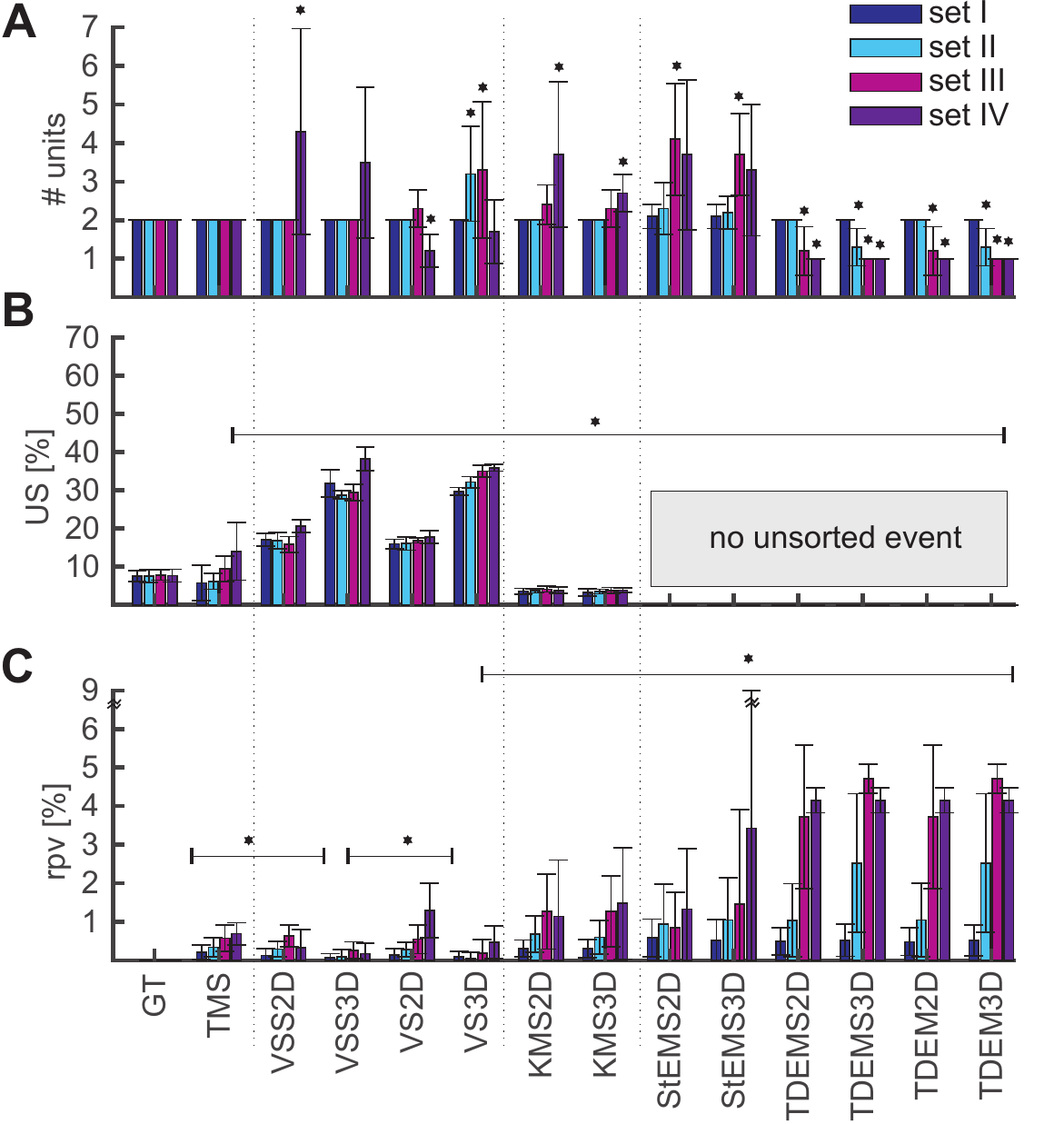}\caption{\textbf{Evaluation of AD sorting results. }\label{figAD} Bar plots
of the sorting results in dependence of the sorting methods (color
coded spike pair similarity in setI to setIV): (A) average number
of detected SUs, (B) percentage of events left unsorted, and (C) percentage
of refractory period violations (rpv). Shown are mean +/- SD of the
10 realizations, stars indicate a significant difference compared
to the GT values (p<0.039 after Bonferroni correction).}
\end{figure}

Fig.$\,$\ref{figAD} presents the first part of the sorting results
obtained for AD sets with varying spike pair similarity. For each
set we again evaluate the resulting number of SUs (Fig.$\,$\ref{figAD}A),
the percentage of US (Fig.$\,$\ref{figAD}B) and rpv (Fig.$\,$\ref{figAD}C).
The GT is shown on the very left of the panels.

For setI and setII (distinct spike pairs), the number of resulting
SUs is mostly quite similar and close to the GT. For setIII and setIV
(similar spike pairs) all TDEM algorithms detect significantly fewer
units, whereas StEMS methods find significantly more units, similar
to KMS and VSS (Fig.$\,$\ref{figAD}A). These observations are similar
to the corresponding ED results (cf. Fig.$\,\ref{figED1}$A). 

EM algorithms do not account for US and KMS methods leave only a very
few US while VS methods yield many more US than present in the GT
(15\% to 40\% compared to 10\%, see Fig.$\,$\ref{figAD}B). The percentage
of US resulting from TMS is mostly close to the GT, only setIV yields
more than 10\% US due to a nearly impossible distinction between pts
and spikes. The major difference to the ED results is the small amount
of US in the GT: Here, US represent artifacts whereas the large amount
of US in the ED are mostly spikes that were left unsorted because
no clear assignment could be made, see Sec.$\,\ref{ResED}$. 

Most methods induce a significant percentage of rpv (Fig.$\,$\ref{figAD}C).
The percentage of rpv is larger the more similar the embedded spike
pairs are, as it is more difficult to separate similar spikes which
induces sorting errors. As observed for the ED, we find that methods
that do not account for US result in a high percentage of rpv, e.g.,
TDEM methods with more than 3\% rpv for setIII and setIV. In contrast,
VS methods yield mostly less than 1\% rpv. 

\begin{figure}
\includegraphics[width=1\columnwidth]{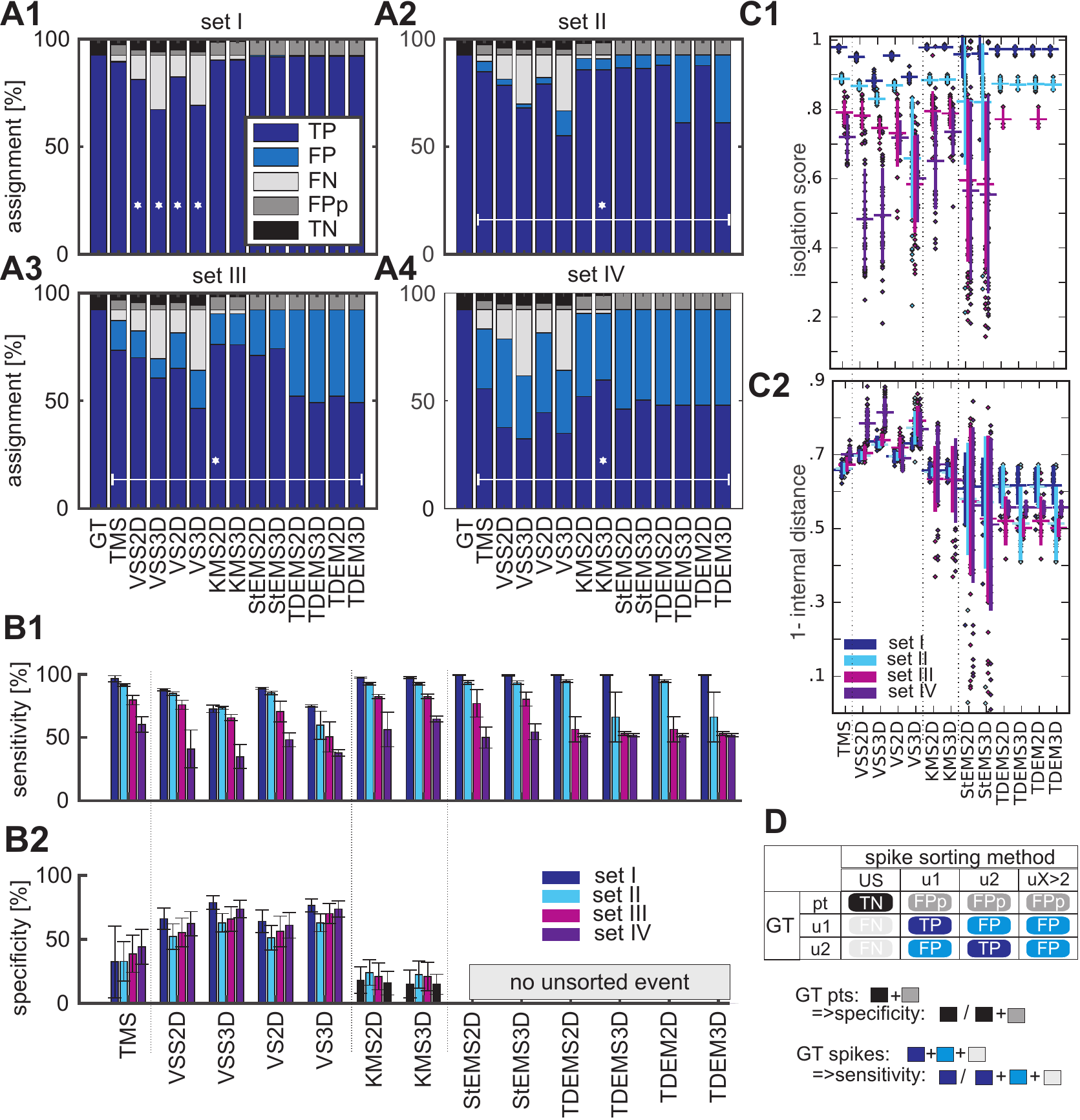} \caption{\textbf{Evaluation of AD sorting results with GT comparison. }\label{figGT}
(A1-A4) Stacked bar plots showing the average percentage of correct
and wrong assignments of spikes and pts for the AD in dependence of
the sorting methods: TP indicates correctly classified spikes, FP
misclassified spikes, FN spikes wrongly left unsorted, FPp indicates
pts wrongly classified as spikes, and TN pts correctly left unsorted
(c.f.$\,$D). Stars indicate a significant difference to the GT values
(p<0.039 after Bonferoni correction). (B1, B2) Sensitivity and specificity
measures in dependence of the sorting methods. Shown are mean +/-
SD of the 10 realizations. (C1, C2) Cluster quality measures IS and
1-Di in dependence of the sorting methods: each dot represents the
value obtained for one SU. Horizontal lines indicate the average over
all SUs in 10 realizations, vertical lines indicate the corresponding
SD. The color code in B1, B2, C1, and C2 represents the spike pair
similarity. (D) Summary of TP, TN, FP and FN notations and definition
of sensitivity and specificity measures.}
\end{figure}

Fig.$\,$\ref{figGT}A1 to A4 show the second part of the AD results:
the TP, TN, FP, FPp and FN assignments made for the four sets. The
100\% correct GT assignment consists of two parts: 90\% TP, i.e.,
correctly classified spikes and 10\% TN, i.e., pts that were correctly
left unsorted. Concerning the spike events in setI, most methods perform
quite well, yielding a TP rate close to 90\%. Only VS(S) methods leave
8\% (2D) to 15\% (3D) of spikes unsorted which results in a comparably
high FN and low TP rate. However, as expected from the ED results,
VS(S) methods also correctly leave most pts unsorted (TN close to
10\%). KMS, StEMS, and TDEMS yield generally high FPp and low FN rates:
many pts are wrongly classified as spikes and only a few or no spikes
are left unsorted. FN, FPp and TN rates change only slightly with
increasing spike pair similarity (setI to setIV) since pts are identical
in all sets. The number of misclassified spikes, however, clearly
increases with increasing spike pair similarity: In setII, TDEM(S)3D
already show 30\% FP due to collapsing the spikes from two SUs into
one SU while all other methods yield 2\% to 6\% FP (Fig.$\,$\ref{figGT}A2).
For setIII, all TDEM methods yield approx.$\,$45\% FP while most
other methods result in fewer FP (5\% to 15\%) and correspondingly
higher TP rates (60\% to 80\%). Only VS3D yields less than 50\% TP
due to the relatively high percentage of 20\% FN. 

Fig.$\,$\ref{figGT}C1 and Fig.$\,$\ref{figGT}C2 show that the
cluster quality measures IS and Di mostly reflect the results obtained
by the above GT comparison: the more similar the spike pairs, the
lower the TP rate and the average IS. This agreement holds only partially
for the Di. If the number of identified SUs is large, the resulting
clusters are small and naturally have a small internal distance, e.g.,
the large 1-Di values for VSS in setIV (Fig.$\,$\ref{figGT}C2, cf.
Fig.$\,$\ref{figAD}A). Thus, IS and Di have to be considered in
relation to the number of SUs. VS methods show relatively high 1-Di
values but low TP rates, an effect of the high FN rates which bear
less influence on the Di measure \citep{Joshua2007}. For the TDEM(S)
methods applied to setII, however, the Di results match the TP rates
better than the IS results. 

We expected that using more PCs yields better results. However, the
significant (p<0.05) differences in the percentage of US and TP between
VS(S)2D and VS(S)3D indicate the opposite: the 2D results are closer
to the GT. Still, VS(S)3D yield significantly less rpv compared to
VS(S)2D, but this is simply the consequence of leaving many US. Similarly,
some of the TDEM(S)2D results (\#SU and TP rate for setII) are significantly
closer to the GT than the TDEM(S)3D results. Therefore we conclude,
that VS and TDEM work better in 2D as compared to 3D feature space. 

The differences between the results obtained with and without automatic
scan are inconsistent and only pertain to VS methods. For example,
VSS3D versus VS3D yields mostly significantly (p<0.05) different values
for \#SUs, US and TP where the results obtained with scan are closer
to the GT for \#SU and US but without scan, the TP rates are closer
to the GT. Thus, we see no advantage in applying an automatic parameter
scan. Tab.$\,\ref{tab3}$ and Tab.$\,\ref{tab2}$ in Sec.$\,$\ref{Appendix}
list more quantitative details on the comparative analysis of AD and
ED.

\begin{figure}
\includegraphics[width=0.5\columnwidth]{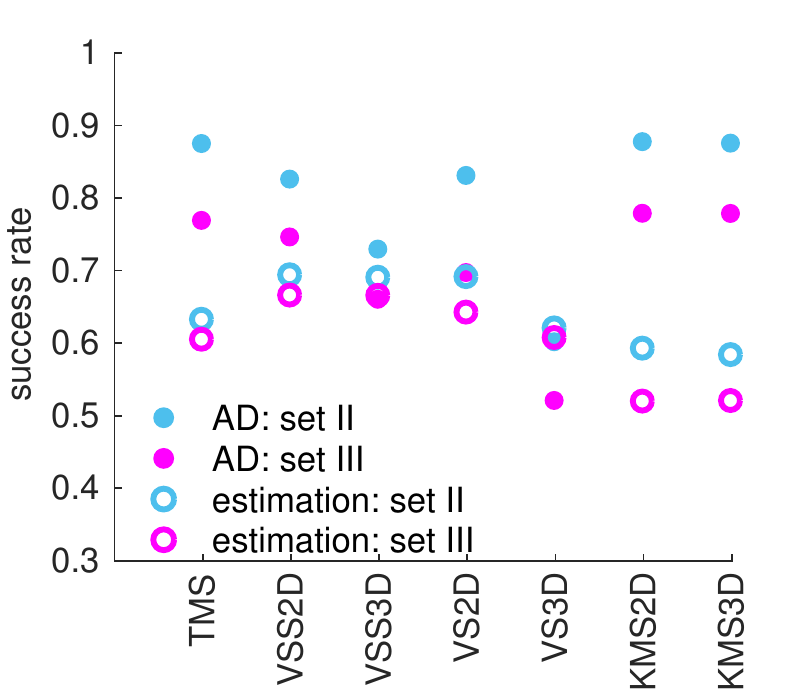}

\caption{\textbf{AD sorting results in dependence of the amount of perturbations\label{figADcompare}.}
Shown are the sum of TP and TN rate of the AD setII and setIII for
the sorting methods that account for US events (GT is at one). Filled
circles indicate the proportion given in the AD, i.e., 90\% spikes
and 10\% pts. Open circles indicate an estimate calculated by re-normalization
assuming 50\% spikes and 50\% pts. The color code represents the spike
pair similarity.}
\end{figure}

Fig.$\,$\ref{figGT}B1 and Fig.$\,$\ref{figGT}B2 summarize our
findings: the sensitivity (normalized TP rate) clearly decreases with
increasing spike similarity, independently of the sorting method:
the more similar a spike pair, the harder is the task to distinguish
the spikes and to sort them into different units. For setIV, all sensitivity
values are close to 50\% indicating that the sorting task is so difficult
that the success rates are bound to be close to chance level. However,
there is no clear dependency of the specificity (normalized TN rate)
on the task difficulty. Since EM methods do not account for US, their
specificity is zero. As expected from the ED results, VS methods show
a high specificity but their sensitivity is rather low. In contrast,
KMS and TMS show again, as observed for the ED, a low specificity
while their sensitivity is relatively high. For the AD, we conclude
that KMS and VS(S)2D yield the best compromise between high sensitivity,
i.e., many correctly classified spikes, and high specificity, i.e.,
many identified pts. Hence, the doubts about VSS3D being the best
sorting method for the ED are justified. Fig.$\,$\ref{figADcompare}
shows the so-called success rate, i.e., the sum of TP and TN rates
(filled circles) for AD sets II and III with 90\% spikes and 10\%
pts. It shows that TMS and KMS are the most successful methods, followed
by VS2D. The open circles are an estimate obtained via re-normalization
with changed proportions for spikes (50\%) and pts (50\%). In this
case VS methods show a higher success rate than KMS and TMS. Thus,
the best method to sort the data depends on the amount of perturbations
(i.e., artifacts) which are to be left unsorted. 

\section{Discussion\label{discus}}

The classification of multi unit activity into SUs is an important
prerequisite for many types of data analysis, e.g., neuronal correlations,
spike-LFP phase coupling, or tuning properties of single cells. The
sorting evaluation procedure described in this study is generally
applicable. We provide a comparative analysis that depicts and characterizes
the differences in the results of a selected set of sorting algorithms,
applied to ED and AD with known GT, respectively. Comparing the results
of the ED to the four AD sets we find that the task difficulty in
the ED is most similar to setIII of the AD, i.e., a hard task due
to similar spike shapes.

We evaluate sorting methods provided by the 'Plexon Offline Sorter',
a frequently used software package \citep{Moran2008,Shinomoto2003,Yang2014,Schrock2009,Kelley2018,Lipski2018,Shimamoto2013}.
Aiming for an objective comparison without any user intervention,
we focus on algorithms that either run with a given default parameter
value or in combination with a parameter scan. We additionally use
supervised TMS in order to contrast our results with this widely-used
\citep{Rutishauser2006,Levy2002a,Raz2000,Steigerwald2008} method.
However, the user intervention in such supervised methods is time
consuming, it inherently includes a human bias \citep{Wood2004} and
it typically requires a parameter optimization\citep{Wild2012}. Therefore,
we aim to identify the most appropriate \textsl{unsupervised} sorting
method. 

In agreement with \citep{Brown2004,Wild2012,Knieling2016} we show
that the results obtained by using different sorting methods differ
significantly, in both ED and AD. There are deviations in the number
of detected SUs, in the percentage of US and rpv, as well as differences
in the cluster quality measures. While the IS is typically used to
select well isolated SUs \citep{Lourens2013,Joshua2007,Deffains2014}
we here apply it to verify an augmented occurrence of well isolated
SUs if the corresponding waveforms are distinct and thus easy to separate.

Most extracellular recordings (in particular STN data, see Sec.$\,$\ref{intro})
contain perturbations, e.g., movement or speech artifacts. Therefore,
it is a clear disadvantage of EM methods that they do not leave any
event unsorted. Even though such perturbations are often removed during
a preprocessing procedure (cf. Sec.$\,$\ref{SS}), a considerable
percentage is typically not identified. For example, approx. 8 out
of 10\% pts survived the preprocessing of our AD. Consequently, the
resulting SUs of all EM methods are contaminated, resulting in high
FPp rates. Among the EM methods, StEM yields the highest sensitivity
and the fewest rpv. Hill et all. \citep{Hill2007,Hill2011} discuss
that the assumption of Gaussian distributions in StEM may be inappropriate
for spike clusters due to spike shapes varying with time. The latter
can be caused by bursting activity which is a prominent feature in
STN recordings \citep{Beurrier1999,Hutchison1998,Steigerwald2008,Chibirova2005}.
Still, StEM works comparably well for our data, possibly due to constant
spike shapes during our short recording duration. 

VS algorithms yield the most specific sorting results, they leave
nearly all pts unsorted. Yet, all VS methods also leave a considerable
amount of spikes unsorted which decreases their sensitivity. For the
AD, only VS(S)2D methods provide a good compromise between specificity
and sensitivity. A previous study \citep{Kretzberg2009} details that
TDEM performs better than VS in clustering artificial data adapted
to resemble extracellular recordings from a turtle's retina. Our AD,
however, explicitly contains perturbations. In such a complex case,
as typical for STN data \citep{Lewicki1998}, the non-parametric approach
taken in the VS(S)2D might provide an advantage because the valleys
separating the SUs do not have to obey a specific parametric form
\citep{Fukunaga1990,Hill2011}. 

KMS is the most sensitive algorithm, only a few spikes are left unsorted
and the amount of missclassifications is acceptable. Yet, it detects
only a very few perturbation and thus has a low specificity. There
is no significant difference between KMS2D and KMS3D. 

At first sight, one expects that more information (i.e., 3D) yields
a better performance but VS and TDEM perform better in 2D feature
space than in 3D. The additional dimension may capture the variability
in the background noise \citep{Lewicki1998,Bishop} and thus lead
to misclassifications. 

Another important point for selecting an appropriate sorting method
is the type of analysis that the user aims to perform with the resulting
SUs. Missed spikes (FN), for example, reduce the significance of spike
synchrony stronger than misclassified spikes (FP) \citep{Pazienti2006}.
Thus, for the analysis of neuronal correlation in STN recordings \citep{Moran2008,Weinberger2006}
KMS is a better choice than VS. Another example are tuning curves,
i.e., the distributions of neuronal firing rates with respect to a
movement\citep{Georgop82} or stimulus \citep{Hubel59} direction.
In this case, misclassified events (spikes, pts) can induce incorrect
multimodal distributions while missed spikes lead to an underestimation
of the true firing rates \citep{Hill2011}. 

Typically, the average firing rates measured in the STN of Parkinson
patients are reported to range from 25$\,$Hz up to 50$\,$Hz \citep{Lourens2013,Steigerwald2008,Deffains2014,Remple2011}.
We observe rates ranging from 14$\,$Hz up to 39$\,$Hz, purely depending
on the sorting algorithm. Thus, we find lower rates than reported
in the literature which can have several reasons: the specific disease
type (tremor dominant versus akinetic-rigid), disease duration \citep{Remple2011},
as well as the exact recording place \citep{Deffains2014}. The method-dependent
dispersion of average firing rate values observed here is 25$\,$Hz
which is identical to the rate dispersion reported in the literature.
Similarly, the amount of regular, irregular, and bursty SUs strongly
depends on the sorting method. Bursting SUs are a characteristic feature
of STN recordings in PD patients \citep{Chibirova2005,Lourens2013,Beurrier1999,Levy2001}
and are reported to vary from 5\% to 25\% \citep{Chibirova2005} or
15\% to 34\% \citep{Lourens2013}, depending on the recording site.
We find a similar amount of variability, namely 7\% to 25\% bursting
SUs, ascribed to the sorting method. 

In summary, different spike sorting approaches yield highly variable
results. In order to recommend a sorting method we distinguish between
two cases: 'clean' and 'noisy' data. With 'clean' we mean that a first
visual inspection of the data indicates that there are only a few
artifacts and distorted spike shapes -- or the given perturbations
can easily be identified and removed otherwise. With 'noisy data'
we mean frequent perturbations that are difficult to identify and
to remove. If the data is relatively clean we recommend to use the
KMS method since it offers the highest success rate (Fig.$\,$\ref{figADcompare})
due to a high sensitivity (Fig.$\,$\ref{figGT}B1) and relatively
well isolated clusters. If the data is particularly noisy and if missed
spikes are less relevant for the subsequent analysis, VS(S)2D is probably
a better choice. It combines a high specificity with an intermediate
sensitivity (Fig.$\,$\ref{figGT}B1,B2) so that its success rate
is higher in case of many aritifacts (Fig.$\,$\ref{figADcompare})
and yields very few rpv. 

The procedure described here could generally serve as a pre-analysis
step to select the appropriate sorting method for a specific data
set: One first generates an AD set with known GT which is adapted
to the experimental recordings. The sorting algorithms in question
are then applied to the AD and the results are evaluated in relation
to the GT. Finally, one selects the method with the best results and
applies it to the experimental recordings. It is elementary enough
to be generally applicable but yields results specific to the given
data. Our results clearly show the importance of a careful spike sorting
method selection.

\section{Appendix\label{Appendix}}

\subsection{Quantitave sorting results analysis }

We here present two additional tables (Tab.$\,\ref{tab2}$ and Tab.$\,\ref{tab3}$)
that quantitatively summarize the sorting results of both ED and AD,
focusing on setII and setIII of the AD as these are most relevant:
the setI spike pair is more distinct and thus easier to distinguish
than ED spikes while the setIV spike pair is so similar and hard to
distinguish that any sorting can only approach chance level, see Fig.$\,\ref{figGT}$A4
and B1. Moreover, the majority of the ED evaluation results are close
to the values obtained for setIII. 

\begin{table}
\caption{\textbf{Evaluation measures for AD and ED. }Summary of evaluation
measures of AD and ED sorting methods: average percentage of refractory
period violations and isolation score for AD and ED. The first two
columns pertain to setII and setIII of the AD. The last two columns
show the average number of units and the average percentage of US
in the ED. \label{tab2}}

\begin{tabular}{|c|c|c||c|c|c|c|}
\hline 
Methods & AD: rpv {[}\%{]}  & IS & ED: rpv {[}\%{]} & IS & \#SU & \%US\tabularnewline
\hline 
TMS & 0.33 | 0.57  & 0.89 | 0.79  & 0.53 & 0.64 & 2.0 & 39.2\tabularnewline
\hline 
VSS 2D & 0.29 | 0.63  & 0.87 | 0.78  & 0.61 & 0.58 & 3.8 & 14.8\tabularnewline
\hline 
VSS 3D & 0.09 | 0.26  & 0.83 | 0.75  & 0.66 & 0.64 & 2.5 & 31.1\tabularnewline
\hline 
VS 2D & 0.28 | 0.55  & 0.87 | 0.73  & 1.07 & 0.66 & 2.4 & 15.0\tabularnewline
\hline 
VS 3D & 0.04 | 0.19  & 0.66 | 0.58  & 0.54 & 0.52 & 3.1 & 43.1\tabularnewline
\hline 
KMS 2D & 0.68 | 1.26  & 0.89 | 0.80 & 1.12 & 0.72 & 2.9 & 2.9\tabularnewline
\hline 
KMS 3D & 0.6 | 1.27  & 0.89 | 0.79  & 1.36 & 0.78 & 2.2 & 2.0\tabularnewline
\hline 
\end{tabular} 
\end{table}
\begin{table}
\caption{\textbf{Evaluation measures for AD.} Summary of AD sorting results
with GT comparison: average number of detected units, percentage of
unsorted events, TN and TP rate. The two entries per column indicate
the results of setII and setIII. \label{tab3}}

\begin{tabular}{|c|c|c|c|c|}
\hline 
Methods & \#SU & \%US  & \%TN & \%TP\tabularnewline
\hline 
\hline 
GT & 2 | 2 & 7.5|7.8 & 7.5| 7.8 & 92.5 | 92.2\tabularnewline
\hline 
TMS & 2 | 2 & 6.2 | 9.4 & 2.6 | 3.2 &  84.7 | 73.5\tabularnewline
\hline 
VSS 2D & 2 | 2  & 16.8 | 15.8 & 4.0 | 4.4 & 78.4 | 70.0\tabularnewline
\hline 
VSS 3D & 2 | 2  & 28.8 | 29.4  & 4.8 | 5.2 & 68.0 | 60.6\tabularnewline
\hline 
VS 2D & 2 | 2.3  & 16.0 | 16.9 & 3.9 | 4.5 & 79.0 | 65.1\tabularnewline
\hline 
VS 3D & 3.2 | 3.3  & 32.1 | 35.0  & 4.8 | 5.5 & 55.2 | 46.4\tabularnewline
\hline 
KMS 2D & 2 | 2.4  & 3.7 | 4.1 & 1.9 | 1.7 & 85.7 | 76.1\tabularnewline
\hline 
KMS 3D & 2 | 2.3  & 3.5 | 3.8 & 1.8 | 1.7 & 85.6 | 76.0\tabularnewline
\hline 
\end{tabular} 
\end{table}

Considering the ED results only, and assuming TMS as GT, VSS3D seems
to be the best sorting method (Tab.$\,\ref{tab2}$): It yields 31\%
US, less than 1\% rvp, a comparably high IS and the \#SUs is close
to the TMS value. However, this rating changes when considering the
AD results. Overall, VS(S)2D yields the best specificity results,
i.e., it leaves most pts unsorted and it yields the lowest amount
of rpv among all supervised methods. The total amount of misclassified
events (FPp+FP) is only 7\% (setII) and 18\% (setIII), see Fig.$\,\ref{figGT},$
but VS methods yield comparable high FN rates: they leave many spikes
unsorted. In contrast, KMS2D yields the best sensitivity results and
it misses only a very few spikes (FN$<$2\%). However, the misclassification
of events (FPp+FP) is higher than in VS(S)2D (10.9\% for setII and
20.4\% for setIII) and leads to a comparably high percentage of rpv.

\section{Acknowledgments}

This work was supported by Deutsche Forschungsgemeinschaft Grant 1753/3-1
Klinische Forschergruppe (KFO219, TP12), Deutsche Forschungsgemeinschaft
Grant GR 1753/4-2 \& DE 2175/2-1 Priority Program (SPP 1665), the
Helmholtz Association through the Helmholtz Portfolio Theme Supercomputing
and Modeling for the Human Brain (SMHB), and by the European Union\textquoteright s
Horizon 2020 research and innovation programme under grant agreement
No. 720270 \& 785907 (Human Brain Project SGA1 \& SGA2). We thank
Paul Chorely and Alexa Riehle for technical help and fruitful discussions.
Tragically, Paul Chorely died before we were able to finish the manuscript. 

\section{References}

\bibliographystyle{abbrv}
\bibliography{spikesortBIB}

\end{document}